\documentclass[useAMS,usenatbib,usegraphicx]{mn2e}
\usepackage{epsf}
\usepackage{amssymb,amsmath}

\title[Consistency testing for invariance of the speed of light at different redshifts]{Consistency testing for invariance of the speed of light at different redshifts: the newest results from strong lensing and type Ia supernovae observations}

\author[Liu et al.]
{Tonghua Liu$^{1,2}$, Shuo Cao$^{2,3\star}$, Marek Biesiada$^{2,4}$ Yuting Liu$^{2}$, Yujie Lian$^{2}$, Yilong Zhang$^{2}$  \\
$^1$ School of Physics and Optoelectronic, Yangtze University, Jingzhou, 434023, China;\\
$^2$ Department of Astronomy, Beijing Normal University, 100875, Beijing, China; \emph{caoshuo@bnu.edu.cn};\\
$^3$ Advanced Institute of Natural Sciences, Beijing Normal
University at Zhuhai 519087, China;\\
$^4$ National Centre for Nuclear Research, Pasteura 7, 02-093
Warsaw, Poland; \emph{Marek.Biesiada@ncbj.gov.p}}

\begin{document}

\date{\today}

\voffset- .5in

\pagerange{\pageref{firstpage}--\pageref{lastpage}} \pubyear{}

\maketitle

\label{firstpage}

\begin{abstract}

The invariance of the speed of light in the distant universe has
profound significance for fundamental physics. In this paper, we
propose a new model-independent method to test the invariance of the
speed of light $c$ at different redshifts by combining the strong
gravitational lensing (SGL) systems and the observations of type-Ia
supernovae (SNe Ia). All the quantities used to test the deviation
of $c$ come from the direct observations, and the absolute
magnitudes of SNe Ia need not to be calibrated. Our results show
that the speed of light in the distant universe is no obvious
deviation from the constant value $c_0$ within the uncertainty based
on current observations. Moreover, we conclude that the currently
compiled SGL and SNe Ia Pantheon samples may achieve much higher
precision $\Delta c/c\sim10^{-2}$ for the deviation of $c$ than all
previously considered approaches. The forthcoming data from the
Legacy Survey of Space and Time and Wide-Field InfraRed Space
Telescope will achieve more stringent testing for deviation of the
SOL (at the level of $\Delta c/c \sim10^{-3}$) by using our
model-independent method. Finally, we discuss the potential ways in
which our technique might be improved, focusing on the treatment of
possible sources of systematic uncertainties.
\end{abstract}


%


   \maketitle
%

\section{Introduction} \label{introduction}

In  modern cosmology, although the $\Lambda$ plus Cold Dark Matter
model ($\Lambda$CDM model) has withstood the test against most of
the observational data, there are still some unresolved issues in
the framework of such standard cosmological model: the Hubble
tension \citep{Freedman17,Valentino2020}, the cosmic curvature
problem \citep{Valentino2019}, and the puzzle of accelerating
expansion of the Universe \citep{Cao11b,Qi18}. Besides the
consideration of possible inconsistencies in different observational
methods, alternative solutions or mechanisms have been proposed to
overcome these problems. One interesting idea is to assume that the
fundamental constants in physics are effective ones, i.e., vary with
cosmic time. Particularly, the constancy of the speed of light
(SOL), which is measured at very high precision in the local
universe ($c_0\equiv2.998\times10^5$ km/s), constitutes one of the
most basic and recognized physical postulates. However, one should
note that the idea of possible deviation from constant speed of
light, originally discussed by Einstein himself \citep{Einstein07},
has been quite debated regarding many theoretical issues
\citep{Dicke57,Petit88,Moffat93,Barrow00,Avelino99,Moffat16}. For
instance, some authors claimed that the time variation of
dimensioned constants such as $c$ are merely human constructs
without any operational meaning \citet{Duff02}, while one could face
serious conceptual problems if the status of $c$ in physics is
changed \citep{Ellis05}. On the other hand, the varying speed of
light (VSL) concepts have also attracted a lot of attention, due to
its extensive applications in investigating fundamental physics and
testing results from standard cosmology. For instance,
\citet{Albrecht99} showed that VSL provides an alternative theory to
the inflationary mechanism, in order to solve the horizon and the
flatness problems in the standard big bang model \citep{Barrow99}.
Such idea was subsequently supported by \citet{Moffat02}, who
pointed out that varying dimensional constants over cosmic history
can have profound implications for the understanding of the laws of
nature, with significant physical consequences directly measured in
experiments and observations. In particular, several observational
analyses have been performed to study possible variation of $c$
focusing on its ability to explain the scale-invariant spectrum of
Gaussian fluctuations in the cosmic microwave background (CMB) data
\citep{Magueijo03}.

In spite of a very high precision of the measurement of $c$ at
present, it should be noted that most of these measurements have
been performed at the Earth's surface (redshift $z=0$). The
observational or experimental data of measuring the SOL in the
distant universe can be rarely seen. However, with the rapid
development of observational technology in modern astronomy, a
variety of observational means and an increasing number of
observational data make it possible to measure the speed of light in
the distant universe with high precision. Quite recently,
 \cite{Salzano15} proposed a method to derive constraints on
the possible variation of $c$ by using Baryon Acoustic Oscillations
(BAO). Their work relies on a simple relation between the maximum of
angular diameter distance $D_A(z_M)$ and the Hubble parameter
function $H(z)$ at the same redshift $z_M$, which can be used to
assess the SOL, i.e., $c(z_M)=D_A(z_M)\cdot H(z_M)$ at a higher
redshift in a simple way. More recently, \citet{Cao17a} performed
the measurement of $c$ at the maximum redshift ($z_M=1.70$) using
(instead of BAO data) the angular diameter distances from
intermediate-luminosity radio quasars calibrated as standard rulers
(the maximum redshift is covered by such observational data set).
Such methodology was subsequently extended by \citet{Salzano17}
focusing on multiple measurements of $c$ at different redshifts, in
which the the degeneracy between the curvature of the universe and
the speed of light was also discussed. Moreover, \citet{Cai16}
developed a new model-independent method to probe the constancy of
$c$ with the elimination of degeneracy between the SOL and the
cosmic curvature. The advantage of their work also lies in the
successful reconstruction of $c$-evolution with redshifts, in a
fully comprehensive way based on Gaussian Processes (GP)
\citep{Seikel12}. However, in their approach one needs the first and
second derivatives of luminosity distances, which cannot be observed
directly but evaluated numerically. Some other observational tests
for varying speed of light in cosmology have also been investigated
in the literature \citep{Qi14,Salzano16,Wang19}.

Fortunately, one new idea for the extragalactic measurement of $c$
dodging some drawbacks of the previous works has been presented by
\cite{Cao20}. The idea is to measure $c$ at multiple different
redshift points by the combination of strong gravitational lensing
systems (SGL) and observations of ultra-compact structure in radio
quasars and its performance has been illustrated on simulated data
from the Large Synoptic Survey Telescope (LSST) of Vera Rubin
Observatory and the very-long-baseline interferometry (VLBI) data on
radio-quasars. Their results showed that the precision of $\Delta
c/c=10^{-4}$ level can be achieved at high redshifts. Inspired by
the above work, we will combine the currently available SGL data and
recent SNe Ia Pantheon sample to perform the measurement of $c$
\citep{Riess98,Perlmutter99}. Although, compared with radio quasars,
the redshift range of Pantheon is inferior to radio quasars, the
greatest advantage of SNe Ia lies in the large sample available.
Therefore, in our work, we use the large sample of real data (SGL
and Pantheon samples) to test the deviation of the SOL over a wide
range of redshift instead of the maximum-condition redshift $z_M$
point. The outline of this paper is as follows. In Section 2 we
present the methodology and the data we use. The results and
analyses are given in Section 3. Finally, we summarize our main
conclusions in Section 4.

\section{Methodology and data}

\subsection{Strong gravitational lensing system}

According to the Einstein's theory of general relativity, the light
will change its route in the gravitational field of massive objects.
When the background light source (at redshift $z_s$), the
intervening galaxy acting as lens (at redshift $z_l$), and the
observer are perfectly aligned or very slightly misaligned, the
source is imaged as the so-called Einstein ring. The size of the
Einstein ring of the source is related to source-lens distance
ratios and the mass distribution model within the lens
\citep{Rusin05,Koopmans06,Koopmans09}. However, it depends on the
assumption that the SOL is constant. On the contrary, for a given
SGL system, if the source/lens distance ratios (for details see
below), the Einstein radius and the mass distribution model are
known, one can perform the measurement of a deviation of the SOL
from the value $c$ measured in the laboratory.

In general, early-type galaxies act as lenses in the majority of SGL
systems detected. Even though their formation and evolution are
still not fully understood in details, a singular isothermal sphere
model (SIS) can reasonably characterize the mass distribution of
massive elliptical galaxies within the effective radius
\citep{Ofek03,Cao16,Liu20}. Under this model, the Einstein radius
can be expressed as
\begin{equation}
\theta_E = 4 \pi \frac{\sigma_{0}^2}{c_{zs}^2}
\frac{D^{A}_{ls}}{D^A_s},
\end{equation}
where $\sigma_0$ is central velocity dispersion of the lens,
${D^A_s}=D^A(0,z_s)$ and ${D^A_{ls}}=D^A(z_l,z_s)$  denote the
angular-diameter distances to the source and between the source and
the lens, respectively. Let us introduce the notation concerning the
speed of light, which will be used in this paper. In the equation
above, $c_{zs}$ is the speed of light $c$ enriched by the subscript
to highlight that this is the quantity to be measured from the
lensing system extending to the source redshift $z_s$. In order to
highlight the speed of light whose value we know from laboratory
measurements we will use the notation $c_0 \equiv c$. It is worth
noting that the measurement of $c_{zs}$ from strong lensing
observations, is derived from geometrical distances related to
space-time metric. Such definition of speed of light which enters
the Hilbert Einstein action in the standard General Relativity
context, is well consistent with that implemented in the previous
works \citep{Salzano15,Cai16}. Following recent analysis of the lens
mass distribution models \citep{Treu06,Cao12,Ma19,Liu19}, a more
general, spherically symmetric power-law mass distribution $\rho
\sim r^{-\gamma}$, where $\rho$ is the total (i.e. luminous plus
dark-matter) mass distribution and $r$ is the spherical radius from
the center of the lensing galaxy has been widely considered in
studies of SGL systems. Assuming that the velocity anisotropy can be
ignored and solving the spherical Jeans equations
\citep{Koopmans06}, one can rescale the dynamical mass inside the
aperture $\theta_{ap}$ projected to the lens plane to the Einstein
radius, and compare it with the mass inside the Einstein radius
$\theta_E$ obtained from strong lensing data. As a result one gets
the following expression:
\begin{equation} \label{Einstein} \theta_E=4
\pi\frac{\sigma_{ap}^2}{c_{zs}^2}\frac{D^{A}_{ls}}{D^A_s}\left(
\frac{\theta_E}{\theta_{ap}} \right)^{2-\gamma} f(\gamma),
\end{equation}
where $f(\gamma)$ is a certain function of the radial mass profile
slope \citep{Cao15,Cao20}, $\sigma_{ap}$ represents the luminosity
averaged line-of-sight velocity dispersion of the lens inside the
aperture radius, and $c_{zs}$ is the speed of light to be determined
from SGL system. This specific symbol $c_{zs}$ emphasizes the
extragalactic measurement of $c$. For a specific SGL system observed
in some survey, one can get the Einstein radius $\theta_E$, velocity
dispersion of the lens inside the aperture radius $\sigma_{ap}$, the
redshift of lens and source $z_l$ and $z_s$. Then, in order to test
the deviation of the SOL, it is necessary to obtain the source-lens
distance ratio $D^{A}_{ls}/{D^A_s}$, which is difficult. Let us note
that in many previous works, the assumption that the SOL is a
fundamental constant of known value, the same reasoning was used to
determine $D^{A}_{ls}/{D^A_s}$ from SGL systems. However, if one
assumes the Friedman-Robertson-Walker (FRW) background metric one
can use the so-called distance sum rule, i.e.,
$D_{ls}=D_s\sqrt{1+\Omega_kD_l^2}-D_l\sqrt{1+\Omega_kD_s^2}$
\citep{Bernstein06,Clarkson2008,Rasanen2015}, where $D$ is the
proper distance related to the angular diameter distance as
$D(z_1,z_2)=D_A(z_1,z_2)(1+z_2)$. Hence, one is able express the
source-lens (angular diameter) distance ratio in terms of angular
diameter distances from lens to observer ($D^A_l$) and from source
to observer ($D^A_s$)
\begin{eqnarray}\nonumber
\frac{D^A_{ls}}{D^A_{s}}&=&\sqrt{1+[(1+z_l)D^{A}_l]^2\Omega_k}\\
&-&
\frac{1+z_l}{1+z_s}\frac{D^A_{l}}{D^A_{s}}\sqrt{1+[(1+z_s)D^{A}_s]^2\Omega_k}.
\end{eqnarray}
One has to keep in mind that our strategy of measuring SOL at
different redshifts is still highly-degenerated with cosmic
curvature, i.e., there might be a possibility that an incorrect
choice of $\Omega_k$ could potentially bias the (otherwise)
model-independent test of the invariance of speed of light. However,
considering the stringent constraints on the spatial curvature
$\Omega_k=0.001\pm0.002$ yielded by the recent Plank CMB
observations together with the BAO data \citep{Planck18}, it is
reasonable to assume the flat Universe. It is beyond the scope of
the present paper to thoroughly discuss the issue of $\Omega_k$ with
the data we use. In the framework of a flat universe, Eq.~(3) could
be rewritten as
\begin{equation}
\frac{D^A_{ls}}{D^A_{s}}=1-\frac{1+z_l}{1+z_s}\frac{D^A_l}{D^A_s}.
\end{equation}
Now the SOL measurement from a specific gravitational lensing system
is obtained as
\begin{equation}
c_{zs}=\sigma_{ap}\sqrt{\frac{4\pi}{\theta_E}\left(1-\frac{1+z_l}{1+z_s}
\frac{D^A_l}{D^A_s}\right)\left( \frac{\theta_E}{\theta_{ap}}
\right)^{2-\gamma} f(\gamma)}.
\end{equation}
In addition to this, another issue which needs clarification is the
central velocity dispersion of the lens, i.e., whether the
measurement of $\sigma_{ap}$ is dependent on the value of SOL. Such
issue was extensively discussed by \citet{Van03}, in which they used
the near ultraviolet lines blue-ward of the CaII, H and K lines for
measuring velocity dispersion, and determined the central velocity
dispersions from a fit to a convolved template star spectrum in real
space. However, the usual procedure to determine velocity
dispersions is to compare a galaxy spectrum with a star spectrum
taken through the same spectrograph with the same setup
\citep{Rix92}. Hence, the spectral lines become wider because of the
Doppler effect and $\sigma_{ap}$ can be roughly inferred from the
observed quantity $c_0\Delta \lambda/\lambda$. It should be noted
that only dimensionless quantities can have invariant, fundamental
meaning, from the perspective of theoretical physics. Therefore, in
this analysis we introduce the quantity of $T_{zs}=c_{zs}/c_0$, with
the following expression of
\begin{equation}
T_{zs}=\frac{\sigma_{ap}}{c_0}\sqrt{\frac{4\pi}{\theta_E}\left(1-\frac{1+z_l}{1+z_s}
\frac{D^A_l}{D^A_s}\right)\left( \frac{\theta_E}{\theta_{ap}}
\right)^{2-\gamma} f(\gamma)},
\end{equation}
Here $T_{zs}$ quantifies the deviation of $c_{zs}$ from $c_0$, with
$T_{zs}=1$ denoting the invariance of speed of light at different
redshifts.

Now one still needs to determine the angular-diameter distances
$D^A_l$ and$D^A_s$ to the lens and to the source. For this purpose,
one could turn to other model-independent astronomical probes, such
as SNe Ia (standard candles), gravitational wave signals (standard
sirens) or compact intermediate-luminosity radio quasars (standard
rulers) to derive the above mentioned distances only according to
the redshifts. In this work, we will use the current newly-compiled
SNe Ia data (Pantheon sample).

\subsection{Strong gravitational lensing sample}

\begin{figure}
\centering
\includegraphics[scale=0.5]{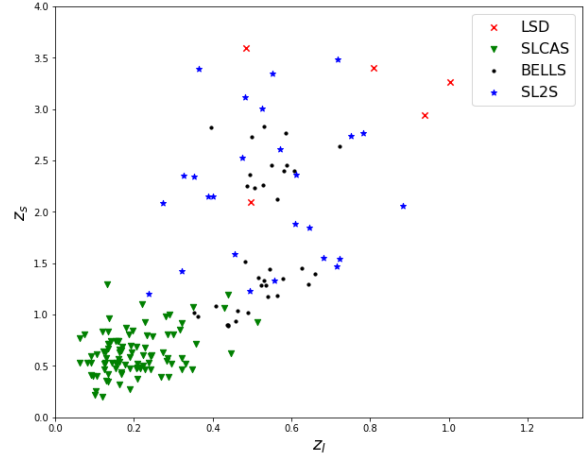}
\caption{The scatter plot of 161 SGL systems from SLACS, BELLS,
SL2S, LSD surveys. }
\end{figure}

Following the work of \citet{Cao15}, we note that the lens galaxies
should ensure the validity of the spherically symmetric hypothesis,
which is guaranteed by the following two selection rules: (i) the
lens galaxies should be of early-type; (ii) the lens galaxies should
not have obvious substructures or close massive companions (not only
physical ones but also projected). Recently, \citet{Chen19} compiled
a sample including 161 galaxy-scale strong lensing systems which
satisfy the requirements mentioned above. This is currently the
largest sample with both high resolution imaging and stellar
dynamical data. The data of this sample came from four surveys, 95
systems from the Sloan Lenses Advanced Camera for Surveys (SLACS)
\citep{Bolton08,Auger09,Auger10}, 38 systems come from an extension
of the SLACS survey known as ``SLACS for the Masses''
\citep{Shu15,Shu17}, 26 systems from the Strong Lensing Legacy
Survey (SL2S)
\citep{Ruff11,Sonnenfeld13a,Sonnenfeld13b,Sonnenfeld15}, 35 from the
BOSS emission line lens survey (BELLS) \citep{Brownstein12}, 14
systems come from the BELLS (GALLERY) GALaxy-Ly$\alpha$ EmitteR
sYstems \citep{Shu16a,Shu16b}, and 5 systems from Lens Structure and
Dynamics (LSD) survey \citep{Koopmans02,Koopmans03,Treu02,Treu04},
in which the complete information (lens redshift $z_l$,  source
redshift $z_s$, Einstein radius $\theta_E$ and velocity dispersion
$\sigma_{ap}$) can be found in Table. A1 of \citet{Chen19}. The lens
and source redshifts cover the range of $0.0625<z_l<1.004$ and
$0.197<z_s<3.595$. The redshift distribution of the lenses and the
sources is shown in Fig. 1, where we can see that most SGL systems
located at low redshift range come from SLACS.

\subsection{Type Ia supernovae observation and Pantheon dataset}

In order to get ${D^A_l}$ and ${D^A_s}$, we focus on SNe Ia referred
to as standard candles of cosmology. The recent SNe Ia sample called
Pantheon has been released by the Pan-STARRS1 (PS1) Medium Deep
Survey, and contains 1048 SNe Ia covering redshift range from 0.01
to 2.3 \citep{Scolnic18}. One of the obvious advantages for using
this sample is the richness of the sample, and its depth in redshift
which means that it extends to higher redshifts than previous data
sets such as Union2.1 or JLA \citep{Betoule14}. The Pantheon
catalogue combines the subset of 279 PS1 SNe Ia
\citep{Rest14,Scolnic14} with useful samples of SNe Ia from SDSS,
SNLS, various low redshift and HST samples \citep{Scolnic18}. In
general, the use of type Ia SNe, whose lightcurve is characterized
by three or four nuisance parameters, involves their optimization
along with the unknown parameters of the cosmological model.

Fortunately, a new approach called BEAMS with Bias Corrections (BBC)
was applied to the Pantheon data set \citep{Kessler17}.  With the
BBC method, \citet{Scolnic18} reported the corrected apparent
magnitude $m_{B,corr}=m_B+\alpha^*\cdot X_1-\beta\cdot
\mathcal{C}+\Delta M$ for all the SNe Ia. Therefore, the observed
distance modulus of SNe Ia is simply reduced to
$\mu_{SN}=m_{B,corr}-M_B$. The scatter plot of 1048 SNe Ia of
apparent B-band magnitude and $1\sigma$ uncertainty from Pantheon
data set is illustrated in Fig.~2. The total uncertainty of the
distance modulus in the Pantheon data set is reduced to the
observational uncertainty of  the corrected apparent magnitude. For
more details one may refer to \citet{Scolnic18}. The luminosity
distance of SNe Ia in Mpc is given by a well-known equation of the
following form
\begin{equation}
D_{L,SN}=10^{0.2\mu_{SN}+1}.
\end{equation}

\begin{figure}
\centering
\includegraphics[width=8cm,height=6cm]{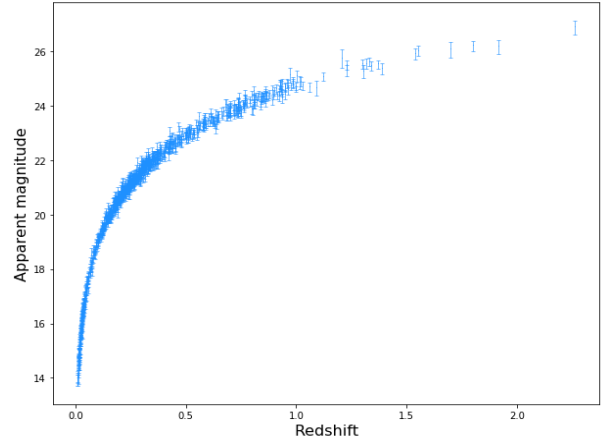}
\caption{The scatter plot of 1048 SNe Ia apparent magnitude and
1$\sigma$ uncertainty from Pantheon dataset.}
\end{figure}

Although the SNe Ia do not provide the angular diameter distance
directly, a fundamental relation called ``distance duality relation"
(DDR) could help us get such distance information
\citep{Etherington1,Etherington2,Cao11,Cao16b}. More precisely, the
luminosity distance and the angular diameter distance at the same
redshift $z$ are connected to each other according to
$D_A(z)=D_L(z)(1+z)^{-2}$. Under the assumption of DDR, the
combination of Eqs.~(6)-(7) will provide the SOL ratio as
\begin{equation}
T_{zs}=\frac{\sigma_{ap}}{c_0}\sqrt{\frac{4\pi}{\theta_E}\left(1-\frac{1+z_s}{1+z_l}
10^{0.2 \Delta m_{B,corr}}\right)f(\gamma)\left(
\frac{\theta_E}{\theta_{ap}} \right)^{2-\gamma} },
\end{equation}
where $\Delta m_{B,corr}=m_{B,corr}(z_l)-m_{B,corr}(z_s)$ is the
corrected apparent magnitude difference between the redshifts of
lenses and sources. It should be noted that the nuisance parameter
$M_B$ is eliminated, which is beneficial in alleviating the
systematics caused by diverse determinations of the absolute
magnitude of this standard candle. One may argue that DDR can be
violated under certain conditions, e.g., the photons emitted by
sources could be affected by various astrophysical mechanisms in the
process of propagation, such as dust extinction or more exotic ones
like axion conversion \citep{Bassett04,Corasaniti06}. However, our
analysis is not affetced much by this. More specifically, the DDR
parameter $\eta=D_L/D_A(1+z)^{-2}$, which appears when the
angular-diameter distance ratio $D_{l}^{A}/D_{s}^{A}$ is related to
the luminosity-distance ratio (from SN Ia) gets cancelled, hence it
does not introduce any significant uncertainty to the final results.
There might still be the issue of redshift dependence of $\eta$
parameter (i.e. cancellation will not be perfect). However, we do
not have any strong evidence for this from other DDR studies in the
literature.

\section{Simulation and Results}

In order to implement the methodology described above, for each SGL
system with lens redshift $z_l$ and source redshift $z_s$ known, the
corrected apparent magnitude $m_{B,corr}(z_l)$ and $m_{B,corr}(z_s)$
should be derived at the same redshift from SNe Ia observations.  In
order to avoid  systematic errors brought by redshift difference
between SGL and SNe Ia, a cosmological model-independent criterion
should be considered: $|z_{SGL}-z_{SN}|<0.005$, where $z_{SGL}$
represents both redshifts of the lens and of the source.

After the redshift selection criterion we obtained the final sample
containing in total 98 systems: 84 systems come from SLACS, 9 come
from BELLS and 5 come from SL2S. Following the  studies of lensing
caused by early-type galaxies \citep{Koopmans09,Treu06,Cao12}, the
spherically symmetric mass distribution ($\gamma=2.01\pm0.03$) was
applied for the lens model, which was obtained from a joint
gravitational lensing and stellar dynamical analysis of a sample of
massive early-type galaxies \citep{Koopmans06}. The individual
measurements of deviation of the SOL based on current observations
of strong-lensing systems and SNe Ia covering redshift range from
$z_s=0.197$ to $z_s=2.34$ are shown in Fig. 3. Uncertainties have
been assessed from the standard uncertainty propagation formula,
based on (uncorrelated) uncertainties of observable quantities
entering the Eq. (8). The uncertainty of $\gamma$ was also
propagated to the total uncertainty budget.

To better describe our results, first, we firstly turn to inverse
variance weighting \citep{Bevington93} to evaluate the deviation of
the SOL, which is often used in meta-analysis to integrate the
results of independent measurements, and our assessment is
$T_{zs}=0.983\pm0.013$ (corresponding to
$c(z_s)=2.950(\pm0.04)\times10^{5}$ km/s) with full sample.
Meanwhile, we have also used a robust non-parametric statistic
\citep{FeigelsonBabu} of calculating the median $Med({T}_{zs})$ and
the corresponding median absolute deviation $MAD({T}_{zs})$. The
result is $Med({T}_{zs})=0.980\pm0.131$ (corresponding to median of
the SOL $Med({c}({z_s}))=2.94(\pm0.393)\times10^{5}$ km/s) for full
sample.
Such approach 
has been extensively applied to quantify the consistency between the
cosmological constant and evolving dark energy in the so-called
Om$(z)$ diagnostic \citep{Zheng16,Zheng18}. Our findings are in
perfect agreement with the results of previous works
\citep{Cai16,Salzano15,Cao17a}. The SOL derived from the current
data on extragalactic sources is consistent with the value
$c_0\equiv2.998\times10^{5}$ km/s within the uncertainty. There is
no obvious evidence to support a deviation in the SOL at distant
Universe, this is the unambiguous conclusion of our work. Let us
compare this result with the work of \citet{Cao20}, where they
combined the SGL systems and compact radio sources quasars to
constrain $c$, and obtained the speed of light value
$c(z_s)=3.005(\pm0.06)\times10^{5}$ km/s. Our results show that
currently compiled SGL systems combined with the SNe Ia Pantheon
sample may achieve constraints with a higher precision of $\Delta
T_{zs}\sim 10^{-2}$ (corresponding to $\Delta c/c\sim10^{-2}$). The
order of magnitude is the same with the previous work of
\citet{Cao20}, but the precision for the SOL measurement has doubled
by using our method.

\begin{table}
\begin{center}
\caption{The deviation $T_{zs}$ and weighted averages of $c$ with
corresponding uncertainties from the full sample and three
sub-samples.}\begin{tabular}{c|c| c } \hline\hline
Sample & Deviation $T_{zs}$ & Weighted averages $c\times 10^5 km/s$  \\
\hline
Full & $0.983\pm0.013$ &$2.950\pm0.04$ \\
\hline
SLACS   & $0.982 \pm0.017$& $2.945\pm0.05$    \\
\hline
BELLS   &$0.904\pm0.097$& $2.712\pm0.29$   \\
\hline
SL2S   &$1.018\pm0.143$& $3.053\pm0.43$   \\
\hline
\end{tabular}

\end{center}
\end{table}

\begin{figure}
\centering
\includegraphics[width=8cm,height=8cm]{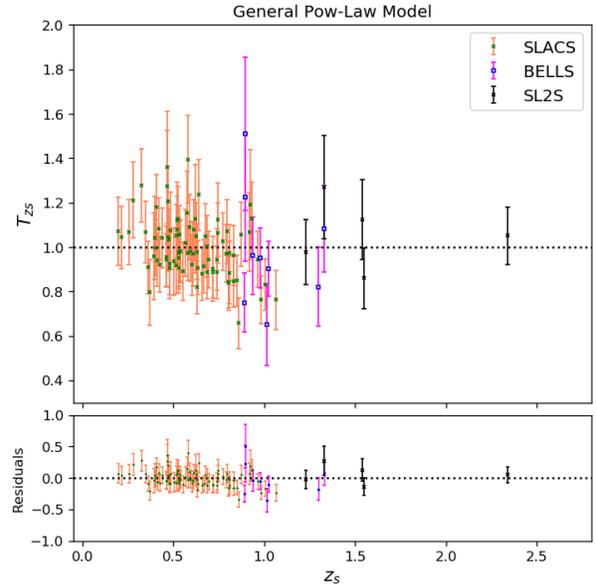}
\caption{ The scatter plot of the deviation $T_{zs}$ of the speed of
light and residuals, based on the current SNe Ia observations and
strong-lensing systems with power-law lens model.}\label{qso_gw}
\end{figure}

On the other hand, a lower central value of the deviation of SOL has
been obtained, i.e., $T_{zs}=0.983<1$. To explore the reason of this
result, we estimated the deviation of the SOL by using SLACS, BELLS
and SL2S surveys sub-samples respectively and summarized them in
Table. 1. One can note that the deviation measurement of the SOL
calculated from SLACS and full sample are almost the same, but the
deviation measurement of the SOL evaluated from SL2S with larger
uncertainty is higher than from full sample. We can, therefore,
infer that the weight of the deviation of SOL assessed from SLACS is
greater than the weight of SL2S, and this will leads to a low the
deviation of SOL. The leverage of SLACS and BELLS systems is
noticeable in Fig.~3, which also illustrates the residual plot for
different lens sub-samples. For SL2S sub-sample, their deviation
$T_{zs}=1.018>1$ can be a result of their systematics and incomplete
samples. Namely, spectroscopic surveys are biased toward smaller
Einstein radii. According to Eq. (6), one can infer that the smaller
Einstein radius and larger velocity dispersion will yield to larger
value for the speed of light, and thus $T_{zs}$ will be greater than
one. Therefore, we took a look at the five SL2S strong lensing
systems (J020524-93023, J022610-042011, J084847-035103,
J084909-041226, and J222148+011542). Their Einstein radii have
smaller values (0''.76, 1''.19, 0''.85, 1''.10, and 1''.40 with the
mean value 1''.06) than Einstein radii from SLACS survey. This is
reasonable, as the source redshift from SL2S is higher than in
SLACS, which means the sources from SL2S is farther away, which
leads to a smaller Einstein radius. It is good to compare this with
the Einstein radius from the whole SL2S sample, because the redshift
matching may cause selection bias. The mean value of the Einstein
radius from 26 SL2S systems yields 1''.32. Similarly, the mean value
of the velocity dispersion of selected five SL2S systems is 255.6
km/s, which is larger than the mean value of 238.9 km/s for the 26
SL2S systems, and further contributes to deviation $T_{zs}>1$. Thus,
we may expect that the insufficient sample size will cause some
deviations in the SOL.

\begin{figure}
\centering
\includegraphics[width=8cm,height=8cm]{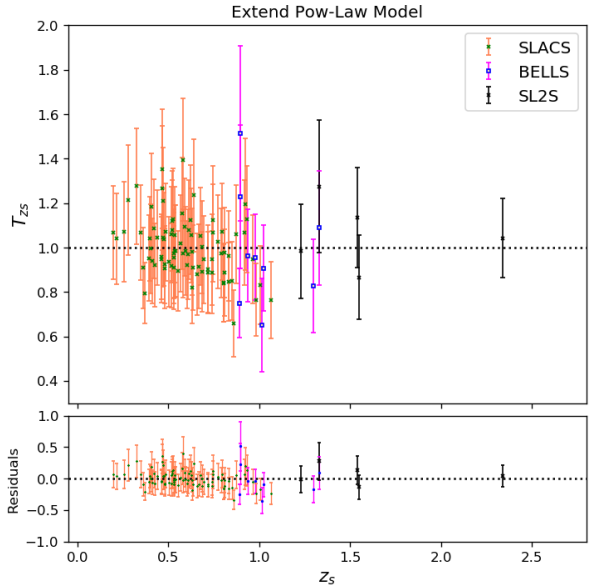}
\caption{The scatter plot of the deviation $T_{zs}$ of the speed of
light and residuals, based on the current SNe Ia observations and
strong-lensing systems with extended power-law (EPL) lens model.}
\end{figure}

There are still several sources of systematics we do not consider in
this paper. For instance, whether the use of a different mass
distribution models for these lenses could significantly affect the
final result. Therefore, we performed a sensitivity analysis and
repeated the above calculation using the extend power-law (EPL) lens
model, in which the luminosity density profile ($v(r)\sim
r^{-\delta}$) is different from the total mass (luminous plus dark
matter) density profile $\rho(r)\sim r^{-\alpha}$. Such lens model
has found widespread astrophysical applications in the literature
\citep{Cao16,Xia17,Qi19}, considering the anisotropic distribution
of stellar velocity dispersion $\beta$
\citep{Koopmans06b,Chen19,Cao17b}. With the EPL lens parameters
$(\alpha, \beta, \delta)$ modeled by Gaussian distributions
$\alpha=2.00\pm0.08$, $\delta=2.40\pm0.11$ and $\beta=0.18\pm0.13$
\citep{Bolton06,Gerhard01,Graur14,Schwab09,Schwab10}, the scatter
plot of the deviation $T_{zs}$ in EPL model is shown in Fig.~4. Our
results provide the deviation $T_{zs}=0.974\pm0.017$ (corresponding
to $c(z_s)=2.922(\pm0.051)\times10^{5}$ km/s), and the median value
$Med({T}_{zs})=0.983$ with the median absolute deviation
$MAD({T}_{zs})=0.259$ for the full lens sample. Therefore, our
results show that the assumed lens model has a slight impact on the
SOL constraint, which highlights the importance of auxiliary data
(such as more high-quality integral field unit) in improving
constraints on the density profile of gravitational lenses. More
detailed models of mass distribution, such as Navarro-Frenk-White
density profile (suitable for dark matter distribution)
\citep{Navarro97}, Sersic-like profile (suitable for stellar light
distribution) \citep{Sersic68}, Peudo-Isothermal Elliptical Mass
Distribution, could also be considered in this context
\citep{Kassiola93}. However, strong lensing observables we used are
determined by the total mass inside the Einstein radius. Hence, they
are not so sensitive to the details of the very central distribution
like cusps (besides the extremal cases). Moreover, the Einstein
rings of the lenses we used corresponded to less than 10 kpc hence
the NFW profile of the dark halo would not likely be manifested.

Fortunately, in the coming years, next generation of wide sky
surveys the LSST survey of the Vera Rubin Observatory, will attain
much improved depth, area, resolution and the sharp increase in the
number of SGL systems is expected \citep{Collett15,Shu18}. The
increase in the number of SGL systems will certainly be very helpful
to alleviate the problems we discussed above. The forthcoming
photometric LSST survey is expected to yield $10^{4 - 5}$ strong
lensing systems \citep{Collett15}, for  which current observational
techniques would allow the redshifts of the lens $z_l$ and the
source $z_s$ to be measured precisely. According to the publicly
released code\footnote{github.com/tcollett/LensPop} with the assumed
LSST survey parameters, as summarized in Table 1 of
\citet{Collett15}, we perform a Monte Carlo simulation of 8000
strong lensing systems with elliptical galaxies acting as lenses,
whose mass distribution is approximated by the singular isothermal
ellipsoid. The strong lensing systems populate the redshift range of
$0.00<z_s\leq3.0$. Concerning the future yields of the LSST, it is
hard to assess the level of precision for identification of fainter,
smaller-separation lenses. Therefore, we applied two selection
criteria: the image separation should be greater than 0.5'', and the
magnitude of lens galaxy in $i$-band should satisfy $m_i<22$. After
the implementation of these two selection criteria, about 3500 lens
systems were retained. For the uncertainty budget, stellar velocity
dispersion $\sigma_v$ and $\theta_E$ from high-resolution images of
the lensing systems are major sources of uncertainty. Following the
recent work of \citet{Liu20}, different fractional uncertainties are
considered for different Einstein radii, i.e., $\theta_E\geq1.5''$
with 3\% uncertainty, $1.0''\leq\theta_E<1.5''$ with 5\%
uncertainty, and $0.5''<\theta_E<1.0''$ with an uncertainty level of
8\%. Based on the fact that the fractional uncertainty of the
velocity dispersion is strongly correlated with the lens surface
brightness, we apply the best-fitted correlation function between
these two quantities in the current SGL sample \citep{Liu20}, in
order to quantify the distribution of the velocity dispersion
uncertainty in the simulated SGL catalogue.

\begin{figure}
\centering
\includegraphics[width=8.5cm,height=6.5cm]{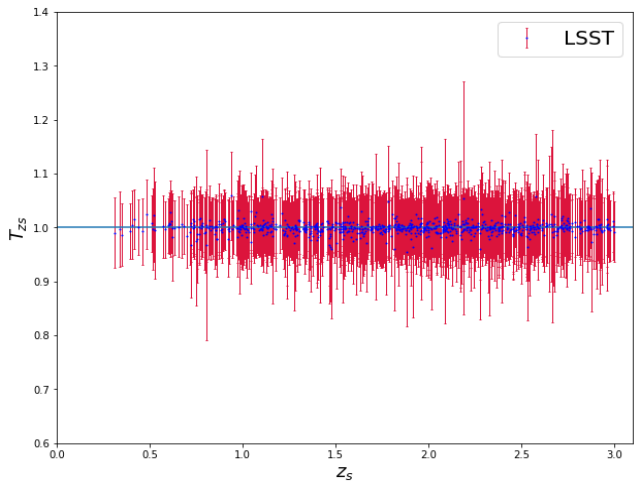}
\caption{Individual measurements of the deviation of the SOL from
the forthcoming LSST survey and WFIRST SNe Ia survey.}\label{future}
\end{figure}

One the other hand, the Wide-Field InfraRed Space Telescope (WFIRST)
is the future NASA mission, which presented a baseline six-year
mission, including a two-year supernova survey strategy,
corresponding to six months of ``on-sky'' time. There are about
$10^{3} - 10^4$ SNe Ia expected to be discovered \citep{Hounsell17}.
In order to create a realistic population of SN Ia sample we perform
a Monte Carlo simulation in the following way: (1) we adopted the
parameterized volumetric rate of SNe Ia as the number density of the
SNe Ia population \citet{Rodney14,Graur14}. In our simulation, a
sample of 3000 SNe Ia covering the redshift range of $0.<z \leq 3.$
has been generated. (2) The peak rest-frame absolute magnitude $M_B$
obeys the gaussian distribution with a mean value of -19.3 mag and
standard deviation of 0.2 mag, while the cross-filter
$K$-corrections were computed from the one-component SNe Ia spectral
template of \citet{Nugent02} (please refer to \cite{Barbary14} for
more details). (3) The total uncertainty budget for the distance
modulus were presented in the Science Definition Team for WFIRST SN
survey, which includes systematic uncertainty
$\sigma_{sys}=0.01(1+z)/1.8$ mag \citep{Hounsell17}, statistical
measurement uncertainties and statistical model uncertainties
$\sigma_{meas}=0.08$ mag \citep{Spergel15}, the lensing uncertainty
modeled as $\sigma_{lens}=0.07\times $ mag \citep{Jonsson10,Holz05},
and intrinsic scatter $\sigma_{int}=0.09$ mag. Combining of the
simulated strong lensing system from LSST and the simulated SN
sample based on WFIRST and applying the redshift match, one can
select 831 strong lensing systems which can be used to measure $c$
using our method. The expected individual results of the deviation
of the SOL are shown in Fig.~5. Our results show that such a
combination of strong lensing systems and SNe Ia will result in more
stringent constraints for the deviation on the SOL at the level of
$\Delta T_{zs}\sim 10^{-3}$ (corresponding to $\Delta c/c \sim
10^{-3}$) in the era of future missions.

\section{Conclusions} \label{conclusion}

In this work, we have proposed a new model-independent method to
explore the deviation of the SOL at high redshift from the $c_0$
value known from laboratory measurements, by combining current
largest SGL and SNe Ia sample. All the quantities that we need are
given directly by observations. For a given SGL system, if the mass
distribution is known, one can directly explore the deviation of the
SOL by combining current SNe Ia observations acting as standard
candle to provide distance information. In order to get the
$D_{ls}/D_{s}$ distance ratio from SNe Ia acting as standard
candles, we have used the assumption of spatial flatness in this
work. However, compared with previous works, our method has the
following advantages:
\begin{itemize}
\item  Exploring the deviation of the SOL over a wide range of redshift instead of focusing on a single value of $z_M$ corresponding to the maximum of $D_A(z)$ function;
\item  Our method need not assume any cosmological model and the samples used here are also model-independent;
\item  The determination of the so called nuisance parameter $M_B$ of SNe Ia will be no longer necessary.
\end{itemize}

First of all, we turn to the recent catalog by \citet{Chen19} that
contains 161 SGL systems with source redshifts ranging from 0.197 to
3.595, obtained in four surveys including SLACS, BELLS, SL2S, and
LSD. Combining these SGL systems together with the estimates of the
angular diameter distances from standard candles (SNe Ia), we obtain
the weighted average value $T_{zs}=0.983$ with the uncertainty
$\Delta T_{zs}=0.013 $ and median value $Med({T}_{zs})=0.980$ with
the median absolute deviation $MAD({T}_{zs})=0.131$ for  full
sample, which indicate that the SOL at high redshifts is fully
consistent with $c_0$ within uncertainty. There is no obvious
evidence to support a deviation in the SOL in a distant Universe.
The analysis of the SLACS sub-sample shows that there is no obvious
deviation from the $c_0$ within 1$\sigma$ confidence region at low
redshift range. For SL2S sub-sample, their deviation $T_{zs}>1$ can
be a result of their systematics and incomplete samples. Thus, we
may expect that the insufficient sample size will cause some
deviations in the SOL. In order to show the full potential for our
method, we also considered the simulated SGL systems from the
forthcoming LSST survey combined with SNe Ia sample from the future
WFIRST survey. Our results show that with these future surveys our
approach of testing the deviation of the SOL may achieve constraints
with much higher precision of $\Delta T_{zs}\sim 10^{-3}$.

In spite of the improved precision for testing the deviation of the
SOL, there are several sources of systematic errors affecting our
results, including the lens modeling, the uses of redshift matching
criterion and the DDR. For the lens modeling, although the lens
system has high resolution imaging at the current technique level,
the uncertainty of stellar kinematic spectral observations of lens
galaxies may lead to 10$\%$ statistical uncertainty in the
measurement of the speed of light. Fortunately, in the coming
optical imaging surveys, one can expect improved  depth, area,
resolution and the sharp increase in the number of SGL systems by
the next generation of wide and deep sky surveys such as the Vera
Rubin Observatory LSST survey \citep{Collett15,Shu18}. High-quality
imaging and spectroscopic observations on SGL systems will certainly
be very helpful to alleviate the problem. For redshift matching
criterion, an interesting idea is to use lensed and unlensed type Ia
supernovae (or other standardized sources) to estimate the speed of
light \citep{Cao18} which can perfectly avoid systematic errors
caused by the redshift match. For the use of the DDR, one can use
alternative probes which directly provide the angular diameter
distances to achieve testing the deviation of the SOL, such as
ultra-compact radio structure acting standard rule \citep{Cao20}.

As a final remark, any possible (statistically significant)
deviation in the SOL might have profound implications for the
understanding of fundamental physics and natural laws. This
encourages us to look forward to the possible method to perform
consistency testing for invariance of the speed of light on
extragalactic objects with higher accuracy and precision in the
future, which strengthens our interest in observational search for
more SGL samples and astronomical probes with smaller statistical
and systematic uncertainties.

\section*{Acknowledgments}

This work was supported by the National Natural Science Foundation
of China under Grant Nos. 12021003, 11690023, 11633001 and
11920101003, the National Key R\&D Program of China (Grant No.
2017YFA0402600), the Beijing Talents Fund of Organization Department
of Beijing Municipal Committee of the CPC, the Strategic Priority
Research Program of the Chinese Academy of Sciences (Grant No.
XDB23000000), the Interdiscipline Research Funds of Beijing Normal
University, and the Opening Project of Key Laboratory of
Computational Astrophysics, National Astronomical Observatories,
Chinese Academy of Sciences. M.B. was supported by the Foreign
Talent Introducing Project and Special Fund Support of Foreign
Knowledge Introducing Project in China. He was supported by the Key
Foreign Expert Program for the Central Universities No. X2018002.

\section*{Data Availability Statements}

The data underlying this article will be shared on reasonable
request to the corresponding author.


\begin{thebibliography}{99}

\bibitem[Albrecht \& Magueijo(1999)]{Albrecht99}Albrecht, A., \& Magueijo, J., 1999,  PRD, 59, 043516
\bibitem[Avelino \& Martins(1999)]{Avelino99} Avelino, P. P., \& Martins, C. J. A. P. 1999, PLB, 459, 468
\bibitem[Auger et al.(2009)]{Auger09}Auger, M. W., et al. 2009, ApJ, 705, 1099
\bibitem[Auger et al.(2010)]{Auger10}Auger, M. W., et al., 2010, ApJ, 724, 511
\bibitem[Barbary(2014)]{Barbary14}Barbary, K. 2014, sncosmo v0.4.2, Zenodo, doi:10.5281/zenodo.11938
\bibitem[Barrow(1999)]{Barrow99} Barrow, J. D. 1999, PRD, 59, 043515
\bibitem[Bassett \& Kunz(2004)]{Bassett04} Bassett, B. A., \& Kunz, M. 2004, ApJ, 607, 661
\bibitem[Bernstein(2006)]{Bernstein06} Bernstein, G. 2006, ApJ, 637, 598
\bibitem[Betoule et al.(2014)]{Betoule14}Betoule, M., et al. 2014, A\&A, 568, 22
\bibitem[Bevington et al.(1993)]{Bevington93} Bevington, P. R., et al. 1993, Computers in Physics 7, 415.
\bibitem[Bolton et al.(2006)]{Bolton06}Bolton, A. S., Rappaport, S., \& Burles, S. 2006, PRD, 74, 061501
\bibitem[Bolton et al.(2008)]{Bolton08}Bolton, A. S., et al., 2008, ApJ, 682, 964
\bibitem[Brownstein et al.(2015)]{Brownstein12} Brownstein, J. R., et al., 2012, ApJ, 744, 41
\bibitem[Cai et al.(2016)]{Cai16} Cai, R. G., Guo, Z. K., \& Yang, T. 2016, JCAP, 08, 016
\bibitem[Cao \& Liang(2011)]{Cao11} Cao, S., \& Liang, N. 2011, RAA, 11, 1199
\bibitem[Cao et al.(2011b)] {Cao11b}  Cao, S., Liang N., \& Zhu, Z.-H. 2011, MNRAS, 416, 1099
\bibitem[Cao et al.(2012)]{Cao12}Cao, S., Pan, Y., Biesiada, M., God{\l}owski, W., \& Zhu, Z.-H. 2012, JCAP, 03, 016
\bibitem[Cao et al.(2015)]{Cao15}Cao, S., Biesiada, M., Gavazzi, R., Pi\'{o}rkowska, A., \& Zhu, Z.-H. 2015, ApJ, 806, 185
\bibitem[Cao et al.(2016a)]{Cao16}Cao, S., Biesiada, M., Yao, M., \& Zhu, Z.-H. 2016a, MNRAS, 461, 2192
\bibitem[Cao et al.(2016b)]{Cao16b}Cao, S., Biesiada, M., Zheng, X., \& Zhu, Z.-H. 2016b, MNRAS, 457, 281
\bibitem[Cao et al.(2017a)]{Cao17a}Cao, S., Biesiada, M., Jackson, J., Zheng, X., \& Zhu, Z.-H. 2017a, JCAP, 02, 012
\bibitem[Cao et al.(2017b)]{Cao17b}Cao, S., Li, X., Biesiada, M., et al. 2017b, ApJ, 835, 92
\bibitem[Cao et al.(2018)]{Cao18} Cao, S., Qi, J., Biesiada, M., et al. 2018, ApJ, 867, 50
\bibitem[Cao et al.(2020)]{Cao20} Cao, S., Qi, J., Biesiada, M., Liu, T., \& Zhu, Z.-H. 2020, ApJL, 888, L25
\bibitem[Chen et al.(2019)]{Chen19} Chen, Y., Li, R., Shu, Y., \& Cao. X. 2019, MNRAS, 488, 3745
\bibitem[Clarkson, Bassett \& Lu(2008)]{Clarkson2008}Clarkson, C., Bassett, B., \& Lu, T. C. 2008, PRL, 101, 011301
\bibitem[Collett(2015)]{Collett15}Collett, T. E. 2015, ApJ, 811, 20
\bibitem[Corasaniti(2006)]{Corasaniti06} Corasaniti, P. S. 2006, MNRAS, 372, 191
\bibitem[Dicke(1957)]{Dicke57} Dicke, R. H. 1957, Reviews of Modern Physics, 29, 363
\bibitem[Duff(2002)]{Duff02} Duff, M.J., 2002 [arXiv:hep-th/0208093 ]
\bibitem[Einstein(1907)]{Einstein07} Einstein, A. 1907, Jahrbuch der Radioaktivitat und Elektronik, 4, 411
\bibitem[Ellis \& Uzan(2005)]{Ellis05} Ellis, G. F. R., \& Uzan, J.-P. 2005, Am. J. Phys. 73, 240
\bibitem[Etherington(1933)]{Etherington1} Etherington, I. M. H. 1933, Phil. Mag, 15, 761
\bibitem[Etherington(2007)]{Etherington2}Etherington, I. M. H. 2007, Gen. Rel. Grav, 39, 1055
\bibitem[Feigelson \& Babu (2012)]{FeigelsonBabu} Feigelson, E. \& Babu G.J., Modern Statistical Methods for Astronomy: With R Applications, Cambridge University Press, Cambridge, 2012
\bibitem[Freedman(2017)]{Freedman17} Freedman W. L. 2017, Nat. Astron, 1, 0169
\bibitem[Gerhard et al.(2001)]{Gerhard01}Gerhard O., Kronawitter A., Saglia R. P., \& Bender R. 2001, AJ, 121, 1936
\bibitem[Graur et al.(2014)]{Graur14}Graur, O., et al. 2014, ApJ, 783, 28
\bibitem[Grillo \& Bertin (2014)]{Grillo08}Grillo, C., Lombardi, M., \& Bertin, G. 2008, A\&A, 477, 397
\bibitem[Holz \& Hughes(2005)]{Holz05}Holz, D. E., \& Hughes, S. A. 2005, ApJ, 629, 15
\bibitem[Hounsell et al.(2017)]{Hounsell17} Hounsell, R., Scolnic, D., Foley, R. J., et al. 2017, arXiv:1702.01747v1
\bibitem[J\"{o}nsson et al.(2010)]{Jonsson10}J\"{o}nsson, J., Sullivan, M., Hook, I., et al. 2010, MNRAS, 405, 535
\bibitem[Kassiola \& Kovner(1993)]{Kassiola93} Kassiola, A. \& Kovner, I. 1993, ApJ, 417, 450
\bibitem[Kessler \& Scolnic(2017)]{Kessler17} Kessler, R. \& Scolnic, D. 2017, ApJ, 836, 56
\bibitem[Koopmans \& Treu(2002)]{Koopmans02}Koopmans, L. V. E. \& Treu, T. 2002, ApJ, 568, 5
\bibitem[Koopmans \& Treu(2003)]{Koopmans03}Koopmans, L. V. E. \& Treu, T. 2003, ApJ, 583, 606
\bibitem[Koopmans(2005)]{Koopmans06b} Koopmans, L. V. E., 2006, EAS Publications Series, 20, 161
\bibitem[Koopmans et al.(2006)]{Koopmans06}Koopmans, L. V. E., Bolton, A. S., Burles, S. and Moustakas, L. A. 2006, ApJ, 649, 599
\bibitem[Koopmans et al.(2009)]{Koopmans09}Koopmans, L. V. E., Bolton, A., Treu, T., et al. 2009, ApJL, 703, L51
\bibitem[Liu et al.(2019)]{Liu19}Liu, T., Cao, S., Zhang, J., et al. 2019, ApJ, 886, 94
\bibitem[Liu et al.(2020)]{Liu20}Liu, T., Cao, S., Zhang, J., et al. 2020, MNRAS, 496, 708
\bibitem[Ma et al.(2019)]{Ma19} Ma, Y. B., Cao, S., Zhang, J., et al. 2019, EPJC, 79, 121
\bibitem[Magueijo(2000)]{Barrow00} Magueijo, J. 2000, PRD, 62, 103521
\bibitem[Magueijo(2003)]{Magueijo03}Magueijo, J. 2003, Rept. Prog. Phys. 66, 2025
\bibitem[Moffat(1993)]{Moffat93} Moffat, J. W. 1993, IJMPD, 2, 351
\bibitem[Moffat(2002)]{Moffat02} Moffat, J. W. 2002 [arXiv:hep-th/0208109]
\bibitem[Moffat(2016)]{Moffat16} Moffat, J. W. 2016, EPJC, 76, 130
\bibitem[Navarro, Frenk \& White(1997)]{Navarro97} Navarro, J. F., Frenk, C. S., \& White, S. D. M. 1997, ApJ, 490, 493
\bibitem[Nugent et al.(2002)]{Nugent02} Nugent, P., Kim, A., \& Perlmutter, S. 2002, PASP, 114, 803
\bibitem[Ofek et al.(2003)]{Ofek03} Ofek, E. O., Rix, H.-W., \& Maoz, D. 2003, MNRAS, 343, 639
\bibitem[Petit(1988)]{Petit88} Petit, J. P. 1988, MPLA, 3, 1527
\bibitem[Perlmutter et al.(1999)]{Perlmutter99} Perlmutter, S., et al. 1999, ApJ, 517, 565
\bibitem[Planck Collaboration(2018)]{Planck18} Aghanim, N., Akrami, Y., Ashdown, M., et al. 2018, arXiv:1807.06209
\bibitem[Qi et al.(2014)]{Qi14} Qi, J., Zhang, M., \& Liu, W. 2014, PRD, 90, 063526
\bibitem[Qi et al.(2018)]{Qi18} Qi, J., Cao, S., Biesiada, M., \& Zhu Z.-H. 2018, RAA, 18, 66
\bibitem[Qi et al.(2019)]{Qi19} Qi, J., et al. 2019, MNRAS, 483, 1
\bibitem[R\"{a}s\"{a}nen, et al.(2015)]{Rasanen2015} R$\ddot{a}$s$\ddot{a}$nen, S., Bolejko, K., \& Finoguenov, A. 2015, PRL, 115, 101301
\bibitem[Rest et al.(2014)]{Rest14} Rest, A., Scolnic, D., Foley, R. J., et al. 2014, ApJ, 795, 44
\bibitem[Riess et al.(1998)]{Riess98} Riess, A.~G., et al. 1998, AJ, 116, 1009
\bibitem[Rix \& White (1992)]{Rix92} Rix, H. W., \& White, S. D. M. 1992, MNRAS, 254, 389
\bibitem[Rodney et al.(2014)]{Rodney14} Rodney, S. A., et al., 2014, AJ, 148, 13
\bibitem[Ruff et al.(2011)]{Ruff11} Ruff, A. J., et al., 2011, ApJ, 727, 96
\bibitem[Rusin \& Kochanek(2005)]{Rusin05} Rusin, D. \& Kochanek, C. S. 2005, ApJ, 623, 666
\bibitem[Salzano, D\c abrowski \& Lazkoz(2015)]{Salzano15} Salzano, V., D\c abrowski, M., \& Lazkoz, R. 2015, PRL, 114, 101304
\bibitem[Salzano, D\c abrowski \& Lazkoz(2016)]{Salzano16} Salzano, V., D\c abrowski, M., \& Lazkoz, R. 2016, PRD, 93, 063521
\bibitem[Salzano(2017)]{Salzano17} Salzano, V. 2017, PRD, 95, 084035
\bibitem[Schwab et al.(2009)]{Schwab09} Schwab J., Bolton A. S., Rappaport S. A., 2009, ApJ, 708, 750
\bibitem[Schwab et al.(2010)]{Schwab10} Schwab, J., Bolton, A. S., \& Rappaport, S. A. 2010, ApJ, 708, 750
\bibitem[Scolnic et al.(2014)]{Scolnic14} Scolnic, D., Rest, A., Riess, A., et al. 2014, ApJ, 795, 45
\bibitem[Scolnic et al.(2018)]{Scolnic18} Scolnic, D., et al. 2018, ApJ, 859, 101
\bibitem[Seikel, Clarkson \& Smith(2012)]{Seikel12}Seikel, M., Clarkson, C., \& Smith, M. 2012, JCAP, 06, 036
\bibitem[S\'{e}rsic (1968)]{Sersic68}S\'{e}rsic, J. L., 1968, \textit{Atlas de galaxias australes} Cordoba, Argentina: Observatorio Astronomico
\bibitem[Shu et al.(2015)]{Shu15}Shu, Y., et al. 2015, ApJ, 803, 71
\bibitem[Shu et al.(2016a)]{Shu16a}Shu, Y., et al. 2016a, ApJ, 824, 86
\bibitem[Shu et al.(2016b)]{Shu16b}Shu, Y., et al. 2016b, ApJ, 833, 264
\bibitem[Shu et al.(2017)]{Shu17}Shu, Y., et al. 2017, ApJ, 851, 48
\bibitem[Shu et al.(2018)]{Shu18}Shu, Y., et al. 2018, ApJ, 864,91
\bibitem[Sonnenfeld et al.(2013a)]{Sonnenfeld13a}Sonnenfeld, A., et al. 2013a, ApJ, 777, 98
\bibitem[Sonnenfeld et al.(2013b)]{Sonnenfeld13b}Sonnenfeld, A., et al. 2013b, ApJ, 777, 97
\bibitem[Sonnenfeld et al.(2015)]{Sonnenfeld15}Sonnenfeld, A., et al. 2015, ApJ, 800, 94
\bibitem[Spergel et al.(2015)]{Spergel15} Spergel, D., Gehrels, N., Baltay, C., et al. 2015, arXiv:1503.03757
\bibitem[Treu \& Koopmans(2002)]{Treu02}Treu, T. and Koopmans, L. V. E. 2002, ApJ, 575, 87.
\bibitem[Treu \& Koopmans(2004)]{Treu04}Treu, T. and Koopmans, L. V. E. 2004, ApJ, 611, 739.
\bibitem[Treu et al.(2006)]{Treu06}Treu, T., Koopmans, L. V. E., Bolton, A. S., Burles, S., \& Moustakas, L. A. 2006, ApJ, 650, 1219
\bibitem[Valentino, Anchordoqui \& Akarsu (2020)]{Valentino2020}Valentino, E. D., Anchordoqui and L. A., Akarsu, O. 2020, arXiv:2008.11284
\bibitem[Valentino, Melchiorri \& Silk(2019)]{Valentino2019} Valentino, E. D., Melchiorri, A., \& Silk, J. 2019, Nat. Astron, 4, 196
\bibitem[van Dokkum \& Stanford(2003)]{Van03}van Dokkum, P. G., \& Stanford, S. A. 2003, ApJ, 585, 78
\bibitem[Wang et al.(2019)]{Wang19} Wang, D. D., Zhang, H. Y., Zheng, J, L., et al. 2019, arXiv:1904.04041
\bibitem[Xia et al.(2017)]{Xia17} Xia, J.-Q., et al. 2017, ApJ, 834, 75
\bibitem[Zheng et al.(2016)]{Zheng16} Zheng, X. G., Ding, X. H., Biesiada, M., Cao, S., \& Zhu, Z. H. 2016, ApJ, 825, 17
\bibitem[Zheng et al.(2018)]{Zheng18} Zheng, X. G., Biesiada, M., Ding, X. H., Cao, S., Zhang, S. X., \& Zhu, Z. H. 2018, EPJC, 78, 274






\end{thebibliography}
\end{document}